\documentstyle[aps,epsf,amsfonts,amssymb,eqsecnum]{revtex} 

\def\ben{\begin{equation}}
\def\een{\end{equation}}
\def\bena{\begin{eqnarray}}
\def\eena{\end{eqnarray}}
\def\ppp{\partial}

\newcommand{\non}{\nonumber}
 
\newcommand{\svs}{\vspace*{3mm}}  
\newcommand{\vs}{\vspace*{5mm}}  

\begin{document}

\def\abstract#1
{\begin{center}{\large \bf Abstract}\par #1 \end{center}}
\def\title#1{\begin{center}{{\Large #1}}\end{center}}
\def\author#1{\begin{center}{{\large  #1}}\end{center}}
\def\address#1{\begin{center}{\it #1}\end{center}} 


\begin{flushright}    
YITP-01-70 \\        
hep-th/0110286 \\
\end{flushright}      
\vskip 0.3cm

\title{\bf Brane Big-Bang Brought by Bulk Bubble } 
\vspace{3mm}

\author{Uchida Gen, Akihiro Ishibashi, and Takahiro Tanaka} 
\vspace{3mm}

\address{
Yukawa Institute for Theoretical Physics, Kyoto University, 
     Kyoto 606-8502, Japan
}        
\vspace{8mm}

\begin{center}
{\large \bf Abstract}
\end{center}
\begin{center}
\begin{minipage}{14cm}
{\small 
 We propose an alternative inflationary universe scenario 
in the context of Randall-Sundrum braneworld cosmology. 
In this new scenario the existence of extra-dimension(s) plays 
an essential role. 
First, the brane universe is initially in the inflationary phase driven by 
the effective cosmological constant induced by small mismatch 
between the vacuum energy in the 5-dimensional bulk and the brane 
tension. This mismatch arises since the bulk is initially 
in a false vacuum. Then, the false vacuum decay occurs, nucleating 
a true vacuum bubble with negative energy inside the bulk. 
The nucleated bubble expands in the bulk and consequently hits the
brane, bringing a hot big-bang brane universe of
the Randall-Sundrum type.
Here, the termination of the inflationary phase is due to 
the change of the bulk vacuum energy. 
The bubble kinetic energy heats up the universe.
As a simple realization, we propose a model, in 
which we assume an interaction 
between the brane and the bubble.  
We derive the constraints on the model parameters 
taking into account the following requirements:        
solving the flatness problem, no force which prohibits 
the bubble from colliding with the brane, 
sufficiently high reheating temperature 
for the standard nucleosynthesis to work, 
and the recovery of Newton's law up to 1mm. 
We find that a fine tuning is needed in order to 
satisfy the first and the second requirements simultaneously,   
although, the other constraints are satisfied in a wide range 
of the model parameters. 
} 
\end{minipage}
\end{center}

\bigskip 


\newpage 

\section{Introduction} 

The braneworld scenario~\cite{Braneworld} provides an intriguing idea that 
our universe exists as a boundary of a higher dimensional bulk spacetime. 
Motivated by the expected progress in the measurement of the Newton's law 
at a smaller length scale, or
inspired by recent progress in M-theory~\cite{HW}, 
various kinds of braneworld models have so far been 
proposed~\cite{Mtheory,LargeExt}.

After a simple but fascinating phenomenological model was 
proposed by Randall and Sundrum~\cite{RS}, 
many studies were done~\cite{Projection,BP,DGPmodel}, 
especially aiming at developing a consistent 
cosmological model in the context of this scenario~\cite{BWC,CP}. 
In the conventional 4-dimensional cosmology, 
the slow-roll inflation is the most favored solution 
to the so-called cosmological problems such as the homogeneity 
problem, the horizons problem, and the flatness problem. 
Rolling of the inflaton field provides a graceful exit of the inflation.  
As far as we know, other alternatives are not so 
successful as the slow-roll inflation
within the context of 4-dimensional theory.  
Hence, on one hand, it is important to examine the extension of this idea to 
the braneworld cosmology~\cite{BraneInflation,Creation}. 
On the other hand, in the braneworld scenario, a new framework for cosmology, 
there may be an alternative which competes with the slow-roll inflation.  
In fact, there have recently appeared interesting 
attempts~\cite{KOST} to solve 
major cosmological problems without the use of inflation. 
Under such a current status of braneworld cosmology,  
it is an important direction of research 
to seek for an alternative scenario in which 
the existence of extra-dimension(s) plays an essential role.

Here we focus on the following fact. 
In 4-dimensional models, the bubble nucleation by the 
first order phase transition 
is less attractive as a mechanism of terminating inflation.  
Because of the causality, the size of 
the nucleated bubble does not exceed the horizon scale.  
Therefore, many bubbles need to nucleate in a synchronized manner 
beyond the horizon scale to realize a sufficiently large and 
homogeneous universe. 
As a mechanism for such synchronization, 
an additional field which controls the transition rate is 
introduced 
in the extended inflation scenario~\cite{extended}. 
However, we may be able to avoid such a complication in 
the context of the braneworld, in which the 4-dimensional 
causality is violated~\cite{Causality}. We pursue this possibility.

We consider an inflation on a boundary brane driven by small mismatch between 
the bulk vacuum energy and the brane tension, and 
discuss the possibility that the nucleation of a true vacuum bubble 
becomes a trigger of the big-bang of the brane universe. 
The point is that the bubble nucleation we consider occurs inside the bulk. 
Not only does such a bulk bubble heat up the brane universe 
through the colliding process, but also provides an anti-de Sitter bulk 
of the Randall-Sundrum setup, 
reducing the effective cosmological constant on the brane 
to zero simultaneously.

For a simple illustration of our idea, we assume that 
5-dimensional universe has nucleated with a 
single positive tension brane 
at the fixed point of ${\mathbb Z}_2$-symmetry~\cite{Creation}. 
The mismatch between the bulk vacuum energy and the brane tension  
drives an inflation on the brane. 
This inflationary phase would last forever if there were 
no mechanism to terminate it. 
However, if the bulk is initially in a false vacuum state, 
the decay of the false vacuum via quantum tunneling results 
in the nucleation of a true vacuum bubble, 
which can be a mechanism to terminate 
the inflation on the brane. 
If the transition occurs with the highest symmetry, 
the nucleated bubble has the common center which respects 
the symmetry of the bulk-brane system. 
However, even if the transition with the highest symmetry is 
the most probable process, quantum fluctuations lead to 
displacement of the position of  
the nucleation from the center of the symmetry. 
In this case, the expanding bubble hits the inflationary 
brane universe. 
As we shall see, 
the intersection of the brane and the bubble wall is spacelike.  
Therefore the energy of the bubble wall is inevitably converted to 
the radiation energy on the brane unless it dissipates into the bulk. 
It is also important to note that 
the brane is instantaneously heated up at this spacelike intersection 
beyond the horizon scale of the brane universe. 
Although such a type of thermalization appears a causality violation 
from the viewpoint of the observers on the brane, 
it is a natural consequence of the bubble nucleation 
in the bulk (outside the brane). 
We call this spacelike intersection a ``big-bang surface.''

In the future of this big-bang surface, the brane evolves as a radiation  
dominated Friedmann-Lemaitre-Robertson-Walker~(FLRW) brane universe. 
The effective cosmological constant is reduced 
with the true vacuum energy chosen to be the negative value which 
balances the tension of the brane. 
Then, the bulk around the brane becomes anti-de Sitter spacetime. 
The gravity is effectively localized on the brane by 
the Randall-Sundrum mechanism. 
Since the true vacuum energy is lower than that in the false vacuum, 
this model allows a creation of anti-de Sitter bulk from de Sitter or 
Minkowski-bulk~\cite{CD}. 
The models to consider bubble nucleation in the bulk have been discussed 
in several different contexts of the braneworld~\cite{bubblenucleation}. 
In particular, a similar idea to realize the Randall-Sundrum setup 
by a collision of bubbles was discussed in Ref.~\cite{Gorsky}, 
where the bubble nucleates through the Schwinger process 
in some external field. Our scenario has similarities with 
the one proposed by Bucher~\cite{Bucher}, in the sense that 
anti-de Sitter bubbles appear as a result of a false vacuum
decay~\cite{CD} and that the hot big bang universe is created by 
a collision of branes. 
The new feature of our scenario is that 
the whole universe begins with a compact spatial section. 
This might be an advantage of our new scenario 
because the well-known proposal of the creation of universe from 
nothing~\cite{HHV} provides a mechanism to explain such an initial state.
We will discuss the relation between these two scenarios in the 
concluding section, too. 

As we shall see, the spatial curvature of the resultant FLRW brane
universe depends on the location of the bubble nucleation in the
bulk. The closer the center of the nucleated bubble is to the center of
the symmetry of the bulk-brane system, the smaller the spatial curvature
of the FLRW brane universe is.
Hence, if the bubble is strongly favored to nucleate near the 
center of the symmetry, the above scenario can be an 
alternative to the standard slow-roll inflation.

This paper is organized as follows. 
In Sec.~\ref{Sect:geometry}, we begin with explaining the dynamics 
of the system that consists of a single de Sitter brane 
at the boundary of the bulk 
and a single anti-de Sitter bubble inside the bulk. 
There we discuss the case that the bulk is given by 
a maximally symmetric 5-dimensional spacetime.  
We treat both the bubble wall and the brane as hypersurfaces. 
We illustrate the geometry of our setup by giving its embedding
into a 6-dimensional flat spacetime. 
In particular, we explain the way how the bubble wall intersects 
the de Sitter brane. 
Then, we discuss the initial condition for the evolution of the brane 
after the brane-bubble collision. 
In Sec.~\ref{Sect:BubbleNucleation}, 
we address the nucleation of a bulk bubble which realizes a
sufficiently flat brane universe.
We discuss this issue by considering a model of 
a bulk scalar field, which has a potential localized 
on the boundary brane as well as that in the bulk, 
with the background bulk geometry fixed.  
We develop a method to treat a quantum tunneling 
on a space with boundary. 
We first consider the process of quantum tunneling described by 
an $O(5)$-symmetric instanton in which 
the bubble center coincides with that of the boundary brane. 
Next, perturbing this instanton, we derive 
the probability distribution of the bubble nucleation 
as a function of the displacement of the center of 
the bubble from that of the bulk-brane system. 
Then, we derive the condition for solving the flatness problem. 
In Sec.~\ref{Sect:Constraint}, we discuss the constraints on 
the model parameters in our scenario 
taking into account the following requirements:
solving the flatness of our universe, 
no force which prohibits 
the bubble from colliding with the brane, 
sufficiently high reheating temperature 
for the standard nucleosynthesis to work  
and recovery of Newton's law up to 1mm. 
We show that most of these constraints are satisfied in a wide range 
of the model parameters. 
However, it turns out to be difficult to 
satisfy the first and the second constraints simultaneously.  
Namely, if we introduce 
stronger interaction between the brane and the bubble to confine 
the bubble nucleation very close to the center of symmetry, 
it becomes harder for the nucleated bubble to hit the inflating brane. 
As a result, we find that our model requires one tuning to adjust  
the strength of interaction. 
Section~\ref{Sect:Conclusion} is devoted to summary and discussion.

\section{Geometry of AdS-bubble and brane universe} 
\label{Sect:geometry}

\subsection{Collision of inflationary brane and AdS-bubble } 
\label{SubSect:collision} 

\begin{figure} 
\centerline{\epsfxsize = 6cm \epsfbox{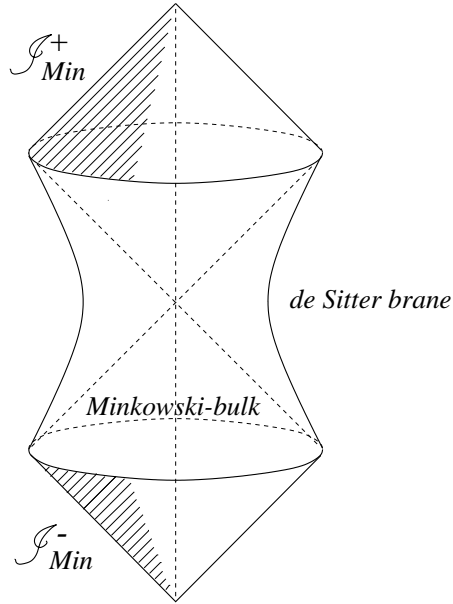}}
\vspace{3mm}  
\begin{center}
\begin{minipage}{14cm}
        \caption{\small  
                  Conformal diagram of a de Sitter brane in Minkowski bulk 
                  ($\ell_F^{-2} \rightarrow 0$ case), 
                  which describes an inflationary brane universe. 
                  A portion of Minkowski spacetime bounded by a hyperboloid,  
         $- (X^0)^2 + (X^1)^2 + (X^2)^2 + (X^3)^2 + (X^{4})^2 = \alpha_B^2$,  
                  is glued along the hyperboloid onto 
                  a copy of itself so that ${\mathbb Z}_2$-symmetry 
                  is satisfied. 
                  The null surfaces ${\cal I}^\pm_{Min}$ denote portions 
                  of the null infinity of Minkowski spacetime. 
               }	\protect \label{fig:dSbrane}
\end{minipage}
\end{center}
\end{figure}

Let us consider the dynamics of a brane $\Sigma_B$ with 
positive tension $\sigma_B$ and the bubble wall $\Sigma_W$ 
with positive tension $\sigma_W$ in the bulk 
given by a 5-dimensional maximally symmetric spacetime, 
i.e., Minkowski, de Sitter or anti-de Sitter spacetime. 
For simplicity, we assume that the brane and the bubble wall 
are infinitesimally thin. It is known that, 
on these assumptions, both the brane $\Sigma_B$ and the world volume 
$\Sigma_W$ are described by $4$-dimensional de Sitter spacetime. 
(See Fig.~\ref{fig:dSbrane})

The relation which determines the respective curvature radii 
$\alpha_B$ and $\alpha_W$ can be obtained as follows. 
Let us consider the case of the bubble wall because 
the result for the brane can be obtained just by equating  
the values of the vacuum energy of the bulk on both sides of the bubble.  
It will be sufficient to consider the case that the both sides 
are anti-de Sitter spacetime. 
We denote the AdS curvature radii on both sides 
by $\ell_F$ and $\ell_T$, respectively. 
The extension to the other cases is almost trivial. 
When the bulk is de Sitter spacetime  
with the curvature radius $\ell_{deS}$, we replace $\ell$ 
in the expressions for the AdS case with $i\ell_{deS}$.  
When the Bulk is Minkowski space, 
we just take the $\ell \rightarrow \infty$ limit. 
The metric of anti-de Sitter spacetime with the curvature radius $\ell$ 
can be written as 
\begin{equation}
  ds^2=dy^2+\ell^2 \sinh^2(y/\ell) d\sigma_{dS_4}^2, 
\end{equation}
where $d\sigma_{dS_4}^2$ is the metric for 4-dimensional 
de Sitter spacetime with unit curvature. 
In this coordinate the vacuum bubble is represented 
by a $y=$constant hypersurface. 
The curvature radius of the bubble wall $\alpha_W$ is related 
to $y_W$, the value of $y$ at the location of the bubble, as 
$\alpha_W=\ell \sinh(y_W/\ell)$.  
The extrinsic curvature on the bubble is easily calculated as 
$K_{\mu}{}^{\nu}=\pm \ell^{-1}\coth(y_W/\ell)\delta_{\mu}{}^{\nu} 
=\pm (\alpha^{-2}+\ell^{-2})^{1/2}\delta_{\mu}{}^{\nu}$. 
The plus (minus) sign corresponds to 
the case in which we choose the side with $y<y_W$ $(y>y_W)$ as the bulk. 
Then we apply Israel's junction condition 
\ben
  K^+_{~\mu}{}^{\nu}  - K^-_{~\mu}{}^{\nu} 
   = \kappa_5 \left( T_{\mu}{}^{\nu} - 
     \frac{1}{3}T\delta_{\mu}{}^{\nu} \right) \,,    
\label{jc:extrinsic-curvature}  
\een
where $T_{\mu}{}^{\nu}$ denotes the intrinsic energy-momentum tensor 
on the hypersurface and $\kappa_5=8\pi G_5$ 
the 5-dimensional gravitational constant. 
For a vacuum bubble, the intrinsic energy-momentum tensor is given by 
$T_{\mu}{}^\nu = - \sigma_W \delta_{\mu}{}^{\nu}$. 
After a simple calculation, we obtain 
\ben 
   \alpha_W^2 := \left[{9\over 4\kappa_5^2 \sigma_W^2}
              (\ell_T^{-2} - \ell_F^{-2})^2 
             - {1\over 2}(\ell_F^{-2} + \ell_T^{-2})+ 
              {\kappa_5^2\over 36} \sigma_W^2\right]^{-1}\,.       
\label{alpha}
\een 
For the brane located at the fixed point 
of ${\mathbb Z}_2$-symmetry, we have 
\begin{equation}  
     \alpha_B^2 := \left[ - \ell_F^{-2} 
               +{\kappa_5^2\over 36} \sigma_B^2\right]^{-1} \,. 
\label{alphaB}      
\end{equation}

To understand the trajectories of the brane and the bubble wall, 
it will be convenient to consider the way how they are embedded in 
the bulk on the false vacuum side. 
As is well-known, anti-de Sitter spacetime is described in 
higher dimensional flat spacetime ${\mathbb E}^{4,2}$ as 
a pseudo-sphere: 
\ben 
  - (X^0)^2 + (X^1)^2 + (X^2)^2 + (X^{3})^2  + (X^{4})^2 
           - (X^5)^2 = -\ell_F^2 \,,  
\label{embed:deSitter}
\een 
where $\{ X^M \}$ are the Cartesian coordinates in ${\mathbb E}^{4,2}$. 
When we consider de Sitter bulk, the same arguments follow by 
taking $X_5$ and $\ell$ as pure imaginary. 
Intrinsic geometry of both the brane and the bubble wall is 
de Sitter space. 
The embedding of 4-dimensional de Sitter spacetime into the  
5-dimensional anti-de Sitter spacetime described above 
is easily done when the center of symmetry of 4-dimensional 
de Sitter spacetime is placed at $X^0=X^1=X^2=X^3=X^4=0$. 
We choose the coordinates so that the de Sitter brane takes this position. 
Then, its embedding is specified with Eq.~(\ref{embed:deSitter}) by 
\ben 
  - (X^0)^2 + (X^1)^2 + (X^2)^2 + (X^3)^2 + (X^4)^2 = \alpha_B^2 \,, 
\label{embed:brane}
\een 
or equivalently 
\begin{equation} 
X^5=\sqrt{\ell_F^2+\alpha_B^2} =: \ell_F \beta_B \,. 
\label{embed:brane2}
\end{equation}

The wall trajectory $\Sigma_W$ can be embedded in the same way. 
However, the centers of symmetry of these two 4-dimensional 
de Sitter surfaces need not be identical. 
The shift of the wall's center can be achieved by (Lorentz) rotation of 
the plane 
\ben 
X^5=\sqrt{\ell_F^2+\alpha_W^2} =: \ell_F \beta_W \,.
\een  
For this purpose, without loss of generality, we can take 
a following Lorentzian rotation in $(X^5,X^1)$-plane 
\bena
    \left( 
       \begin{array}{c}
     	   \bar{X}^5 \\ 
           \bar{X}^1 
       \end{array}
    \right) 
   = 
    \left( 
       \begin{array}{cc}
     	 \cosh \theta_1 & \sinh \theta_1 \\ 
         \sinh \theta_1 & \cosh \theta_1 
       \end{array}
    \right) 
   \left( 
       \begin{array}{c}
     	   {X}^5 \\ 
           {X}^1 
       \end{array}
    \right)   \,, 
\quad 
    \bar{X}^{0,2,3,4} = {X}^{0,2,3,4} \,, 
\label{ct1}
\eena 
and a rotation in $(\bar{X}^5,\bar{X}^0)$-plane 
\bena
    \left( 
       \begin{array}{c}
     	   \bar{\bar X\,}\!{}^5 \\ 
           \bar{\bar X\,}\!{}^0 
       \end{array}
    \right) 
   = 
    \left( 
       \begin{array}{cc}
     	 \cos \theta_0 & - \sin \theta_0 \\ 
         \sin \theta_0 &  \cos \theta_0 
       \end{array}
    \right) 
   \left( 
       \begin{array}{c}
     	   \bar{X}^5 \\ 
           \bar{X}^0 
       \end{array}
    \right)   \,, 
\quad 
    \bar{\bar X\,}\!{}^{1,2,3,4} = \bar{X}^{1,2,3,4} \,. 
\eena 
Then, the wall's trajectory $\Sigma_W$ is written as the intersection of 
the pseudo-sphere~(\ref{embed:deSitter}) and the plane 
\ben
   \bar{\bar X\,}\!^5 = \ell_F \beta_W \,. 
\een 
Since we are concerned with the case that 
the displacement of the center 
$X^M = (- \Delta^0, \Delta^1, 0, 0, 0,\ell_F +O(\Delta^2))$ of $\Sigma_W$ 
from that of the brane $\Sigma_B$ is very small, 
we assume that the boost and the rotation angles, $\theta_1$ and $\theta_0$, 
are so small that they are related to $\Delta$s as 
\ben
    {\Delta^1 \over \ell_F} = \sinh \theta_1  \,, 
\quad  
    {\Delta^0 \over \ell_F} = - \sin \theta_0 \,. 
\een 
We neglect the terms of $O(\Delta^2)$ in the following discussion. 
Then, in terms of the original coordinates $\{ X^M\}$, 
the wall's trajectory can be described as a hypersurface specified by 
\bena
 && - (X^0+ \Delta^0)^2 + (X^1 - \Delta^1)^2 
    + (X^2)^2 + (X^3)^2 + (X^{4})^2 - (X^5 - \ell_F)^2 
    = - 2 \ell_F^2 (1 - \beta_W) \,,  
\label{embed:wall-quad}  
\\     
 && \frac{\Delta^0}{\ell_F} X^0 + \frac{\Delta^1}{\ell_F} X^1 
    + X^5 \approx \ell_F \beta_W \,. 
\label{embed:wall}  
\eena

Note that the bulk of the present system 
has $O(4,2)$-symmetry, but the existence of 
the de Sitter brane partly breaks 
this symmetry. 
The system composed of the bulk and the brane without 
bubble walls has $O(4,1)$-symmetry. 
The nucleation of a bubble wall in this system further 
violates this $O(4,1)$-symmetry unless both 
$\Delta^0$ and $\Delta^1$ are zero.

The embedding of the intersection of the brane $\Sigma_B$ and 
the bubble wall $\Sigma_W$ is given by 
Eqs.~(\ref{embed:deSitter}), (\ref{embed:brane})(or (\ref{embed:brane2})), 
and Eqs.~(\ref{embed:wall-quad}) and (\ref{embed:wall}). 
Since the smaller in absolute value between $\Delta^0$ and $\Delta^1$ can be 
set to zero by a coordinate transformation, the geometry of the intersection 
is classified into the following three cases.

\vs \noindent 
(i) $\Delta^0=0 $ case   

\begin{figure}  
\centerline{\epsfxsize = 9.5  cm \epsfbox{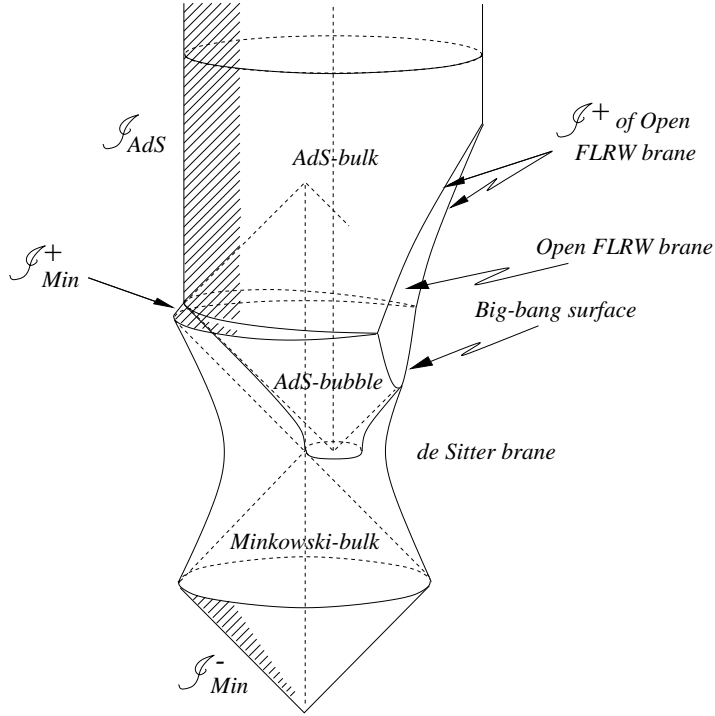}}
\vspace{3mm}
\begin{center}
\begin{minipage}{14cm}
        \caption{\small   
         (i) $\Delta^0= 0, \Delta^1 >0 $ case. 
          The conformal diagram shows the brane big-bang scenario 
          in the case of vanishing false vacuum energy 
          ($\ell_F^{-2} \rightarrow 0$). 
          An AdS-bubble is nucleated and expands in Minkowski spacetime  
          bounded by an inflating de Sitter brane. 
          The two centers of the AdS-bubble and the de Sitter brane are 
          in spacelike separation. The expanding AdS-bubble eventually 
          collides with a portion of the de Sitter brane. 
          The intersection, i.e., the big-bang surface, has a hyperbolic 
          geometry ${\mathbb H}^3$ and an open FLRW brane universe with  
          the AdS-bulk is realized after the brane big-bang.    
          This geometry is glued along the boundary surfaces, 
          except ${\cal I}_{AdS}$ and ${\cal I}^\pm_{Min}$, 
          onto a copy of itself with ${\mathbb Z}_2$-symmetry being satisfied. 
          The timelike surface ${\cal I}_{AdS}$ here denotes the infinity 
          of universal cover of anti-de Sitter spacetime.   
                  }        \protect \label{fig:Openbrane}  
\end{minipage}
\end{center}
\end{figure}

In this case, the center of the bubble is spatially separated 
from that of the brane. 
The intersection $\Sigma_B \cap \Sigma_W$ is specified 
by a simple set of three equations as
$X^1 \approx  ( \beta_B - \beta_W ) \ell_F^2 / \Delta^1$,  
$X^5= \ell_F \beta_B$ and  
\ben
  - (X^0)^2 + (X^2)^2 + (X^3)^2 + (X^4)^2 
 \approx - a_{i}^2  \,, 
\een
where $a_{i}$ is given by 
\ben
  a_{i}^2 := \frac{\ell_F^4}{(\Delta^1 )^2} 
       \left( \beta_B - \beta_W \right)^2 \,.   
\label{RH-radius-1}  
\een
Clearly, this is an $3$-dimensional hyperboloid  
${\mathbb H}^3$ with the curvature radius $a_{i}$. 
As we have anticipated before, 
the intersection, i.e., the big-bang surface, is spacelike. 
Note that as seen from Fig.~\ref{fig:Openbrane}, 
the collision, hence the big-bang, happens only on a part of 
the brane. The remaining part of the brane 
continues the inflation.

\vs \noindent 
(ii)  $\Delta^1=0$ case  

\begin{figure}  
\centerline{\epsfxsize = 5.8cm \epsfbox{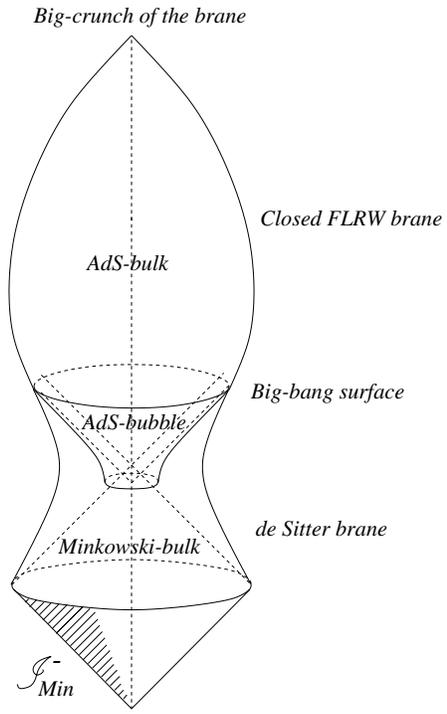}}
\vspace{3mm} 
\begin{center}
\begin{minipage}{14cm}
        \caption{\small    
         (ii) $\Delta^0  >0, \Delta^1 =0$ with vanishing false vacuum energy 
          case. 
          The two centers of the AdS-bubble and the de Sitter brane are 
          in timelike separation. The geometry of the big-bang surface 
          becomes a $3$-sphere ${\mathbb S}^3$ and 
          hence a closed FLRW brane universe with the AdS-bulk 
          is realized after the brane big-bang. The closed FLRW brane finally 
          collapses to a brane big crunch.  
                 }         \protect \label{fig:Closedbrane}
\end{minipage}
\end{center}  
\end{figure}

In this case the separation of 
the centers of the bubble and the brane is timelike. 
To have the intersection $\Sigma_B\cap \Sigma_W$ 
in the expanding side of the de Sitter brane, 
$\Delta^0$ must be positive. 
The intersection is specified by 
$X^0 \approx (\beta_B - \beta_W )\ell_F^2 / \Delta^0$,  
$X^5= \ell_F \beta_B$ and  
\ben
  (X^1)^2 + (X^2)^2 + (X^3)^2 + (X^{4})^2  \approx a_{i}^2 \,.  
\een  
This is an $3$-dimensional sphere ${\mathbb S}^3$ 
with the curvature radius $a_{i}$ given again 
by Eq.~(\ref{RH-radius-1}) with $\Delta^1$ replaced by $\Delta^0$.  
In this case the whole region of the de Sitter brane collides with 
the bubble~(See Fig.~\ref{fig:Closedbrane}). 

\begin{figure} 
\centerline{\epsfxsize = 8.8cm \epsfbox{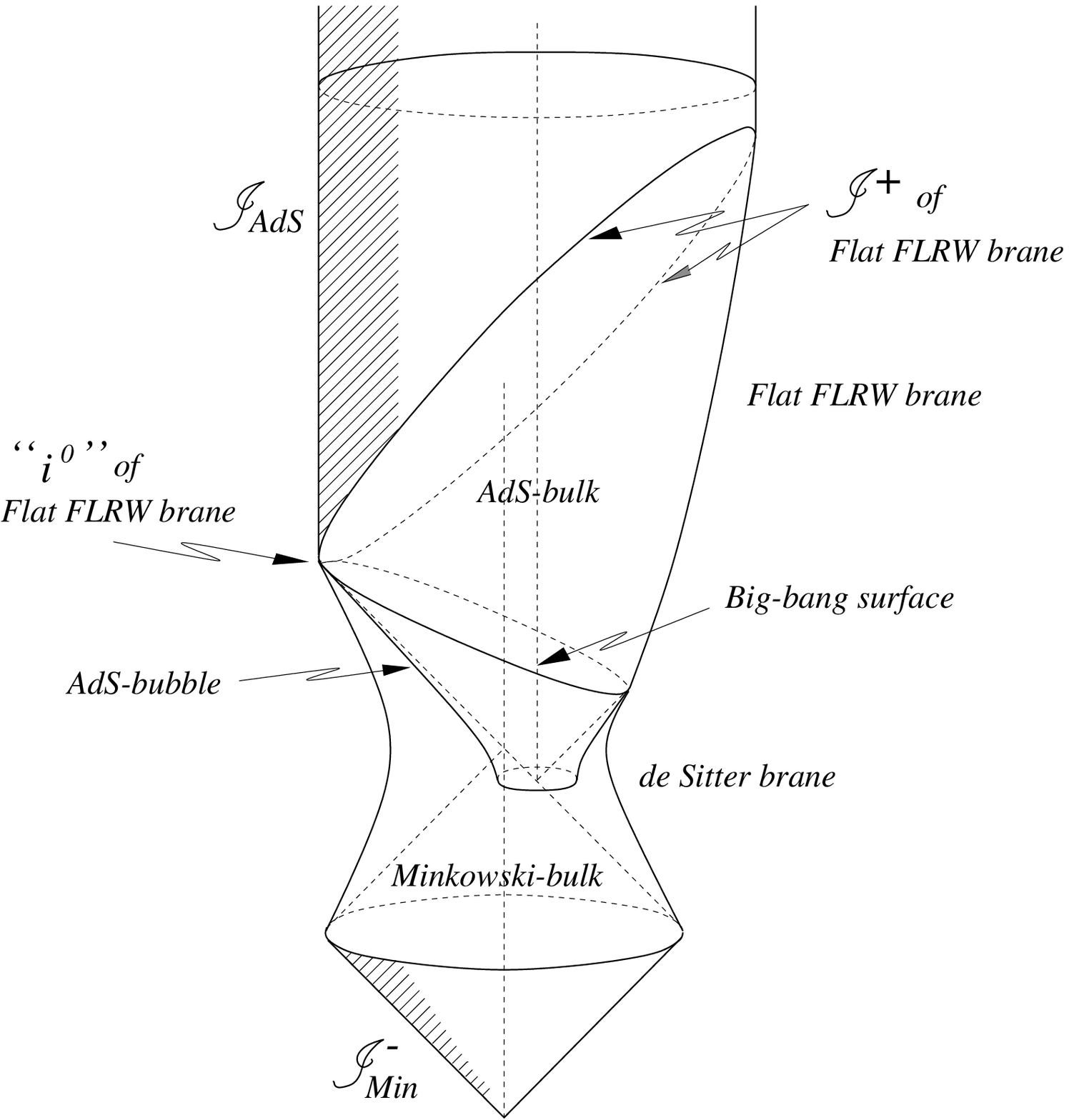}}
\vspace{3mm}
\begin{center}
\begin{minipage}{14cm}
        \caption{\small   
    (iii) $\Delta^0=\Delta^1 >0$ with vanishing false vacuum energy case. 
          The two centers of the AdS-bubble and the de Sitter brane are 
          in lightlike separation. The big-bang surface becomes 
          a flat space ${\mathbb E}^3$, and a flat FLRW brane universe 
          with the AdS-bulk is realized after the brane big bang. 
                 }        \protect \label{fig:Flatbrane}
\end{minipage}
\end{center}  
\end{figure}


\vs \noindent 
(iii) $\Delta^0=\Delta^1$ case 

In this case the two centers of the bubble and the brane are 
in null separation. The case with $\Delta^0=-\Delta^1$ is 
essentially the same. 
The intersection $\Sigma_B\cap\Sigma_W$ is specified by 
$X^5= \ell_F^2\beta_B$, 
\ben  
    X^0 + X^1 = - \frac{\ell_F^2 (\beta_B - \beta_W)}{\Delta^0} \,,   
\label{RH-radius0}
\een 
and Eq.~(\ref{embed:deSitter}). 
Hence, this gives a $3$-dimensional flat space~${\mathbb E}^3$. 
Since flat spacelike section is not a Cauchy surface for de Sitter space, 
there is an eternally inflating region 
on the brane~(See Fig.~\ref{fig:Flatbrane}).

\subsection{Evolution of the brane universe after the brane big-bang} 
\label{SubSect:evolution} 

The region of the inflationary brane $\Sigma_B$ which goes across 
the big-bang surface $\Sigma_B \cap \Sigma_W$ is naturally expected 
to continue to a FLRW brane universe with a metric 
\ben
   ds^2_{(4)} = -d\tau^2 + a^2(\tau) d\sigma^2 _{(3, K)} \,,  
\een 
because the big-bang surface is a $3$-dimensional homogeneous 
and isotropic spacelike surface ${\mathbb H}^3,{\mathbb S}^3, 
or {\mathbb E}^3$ with the metric $d\sigma^2 _{(3, K)}$. 
Here $K$ is the signature of the spatial curvature, which 
is determined depending on the type of collision.  
The spatial sections of the realized FLRW brane universe 
for the collision of type (i), (ii) and (iii), 
are open, closed and flat, respectively. 

As we have mentioned previously, the AdS-bubble nucleation with 
non-vanishing $\Delta$s violates the $O(4,1)$-symmetry 
that the system composed of the brane and the bulk has. 
In the present scenario, we are assuming that the dominant 
process of the bubble nucleation is of the highest symmetry, 
i.e., $\Delta^0=\Delta^1=0$. 
The symmetry breaking is supposed to occur
due to quantum fluctuations in the bubble nucleation process, 
and therefore we have assumed 
that $\Delta$s are sufficiently small compared to the curvature 
radius of the brane. 
The analysis presented above has revealed that 
the curvature scale of the big-bang surface is related to the 
proper distance between the two centers $\Delta^2:= 
-(\Delta^0)^2+(\Delta^1)^2$.  
Under the assumption that $\alpha_W \ll |\ell_F|$, 
we consider two cases, i.e., 
case (a): $\ell_F^{-2}\gg \alpha_B^{-2}$ and 
case (b): $|\ell_F^{-2}|\alt \alpha_B^{-2}$. 
The latter includes the case of the de Sitter 
false vacuum bulk, in which $\ell_F^{2}$ is negative.  
In the respective cases the 
curvature scale of the big-bang surface is estimated as  
\ben 
     a_{i}^2 \approx \frac{\alpha_B^2 \ell_F^2}{|\Delta^2|} \,, 
\quad \mbox{for case (a)}\,;
\quad 
     a_{i}^2 \approx \frac{\alpha_B^4}{|\Delta^2|} \,, 
\quad \mbox{for case (b)}\,. 
\label{aBB=aB/Delta}
\een 
These relations indicate that
$|\Delta^2|$ must be sufficiently small compared to the brane radius 
in order to solve the flatness problem. 
 
After the big-bang, the true vacuum that appears inside 
the bubble provides the AdS-bulk that is a necessary component 
to realize the braneworld scenario of the Randall and Sundrum type. 
The evolution of the FLRW brane universe in the AdS-bulk is determined 
once the composition of the matter fields on the brane is 
specified on the big-bang surface. 
In the present model we simply assume that 
energy of the bubble wall is completely converted into radiation fields 
localized on the brane, 
and the tension of the brane does not change between 
before and after the collision.

The Lorentz factor of the bubble wall seen from the observer 
on the brane is given by (see Appendix\cite{private}) 
\ben
    \gamma \approx \frac{\ell_F^2 
    (\beta_B \beta_W - 1)}{\alpha_B \alpha_W} \,.   
\een 
and is roughly estimated as 
\ben 
 \gamma \approx \frac{\ell_F}{\alpha_W}  \,, 
\quad \mbox{for case (a)} \,;
\quad 
        \approx \frac{1}{2}\frac{\alpha_B}{\alpha_W} \,, 
\quad \mbox{for case (b)}
\een   
Assuming complete conversion of
the energy of the bubble wall, 
the energy density of the radiation field at the onset 
of big-bang on the brane is estimated as 
\ben 
     \rho^{(i)}_r \approx \sigma_W \gamma \,. 
 \label{rhoBB}
\een
Using this estimate of $\rho_r^{(i)}$ and the relation in
Eq.~(\ref{aBB=aB/Delta}), we derive a constraint
on the displacement of the center of the nucleated bubble, $\Delta$. 
Since the spatial curvature term in the FLRW equation 
is not much larger than the contribution from the 
radiation energy density at the present epoch, 
we have $1/a_0^2\lesssim\kappa \rho_r^{(0)}$, or 
$1/a_0^2\lesssim \kappa \rho_r^{(i)} (a_{i}/a_0)^4$. 
Then, these inequalities give
\begin{eqnarray}
   1\lesssim 
    \kappa^2 \rho_r^{(0)} \rho_r^{(i)}a_{i}^4
     \simeq 10^{-128}\rho_r^{(i)}a_{i}^4 \,.
\label{constraint:flatness}
\end{eqnarray}
Substituting Eqs.~(\ref{aBB=aB/Delta}) and (\ref{rhoBB}), we obtain 
\bena 
  { |\Delta| \over \alpha_B}  
      &\lesssim&  
  10^{-32} \left( \frac{\sigma_W \ell_F^5}{\alpha_W} \right)^{1/4}\,, 
\quad \mbox{for case (a)} \,, 
\label{condi:flatness-b-l-w}  
\non 
\\ 
  { |\Delta| \over \alpha_B}  
      &\lesssim&  
  10^{-32} \left( \sigma_W\alpha_B^4\frac{\alpha_B}{\alpha_W}\right)^{1/4}\,, 
\quad \mbox{for case (b)} \,. 
\label{condi:flatness}  
\eena  
The displacement of the center of the nucleated bubble 
must satisfy this inequalities. 
Otherwise, we end up with a curvature-dominant universe, which 
contradicts with observations.

The energy density on the brane 
after the big-bang is given by the sum of the contribution from 
the radiation field and the tension of the brane as 
$\rho_r + \sigma_B$,  
and the bulk is given by anti-de Sitter spacetime 
with the curvature radius $\ell_T$. 
Then, the FLRW equation on the brane becomes
~\cite{BWC}
\ben
 \left(\frac{\dot{a}}{a}\right)^2 
      = \frac{\kappa}{3}\rho_r 
        + \frac{\Lambda_4}{3} - \frac{K}{a^2} 
        + \frac{\kappa_5^2}{36}\rho_r^2 \,, 
\label{eq:Hubble}
\een  
where $\kappa := \kappa_5^2 \sigma_B/6$ denotes 
the effective 4-dimensional gravitational constant on the brane, 
and 
\ben
 \Lambda_4 :=  \frac{{\kappa_5}^2 }{12} \sigma_B^2 - \frac{3}{\ell_T^2} \,, 
\label{4Lambda}
\een
is the effective cosmological constant on the brane.   
After the big-bang, the effective 
cosmological constant $\Lambda_4$
is supposed to vanish. Therefore we need to set  
\ben
 \ell_T^{-2}=(\kappa_5^2/36)\sigma_B^2. 
\label{ellT}
\een 
Although we need here the fine-tuning of the model parameters, 
it is nothing but the usual problem of tuning the cosmological constant. 
It will be interesting to notice the following point. 
Since the 5-dimensional vacuum energy is lower on the true vacuum side, 
we have $\ell_T^2<\ell_F^2$. This inequality is consistent 
with our scenario that the effective cosmological constant on 
the brane was positive before the brane big-bang bubble.

\section{Bubble nucleation} 
\label{Sect:BubbleNucleation}       
In the preceding section, we found that
the constraint~(\ref{condi:flatness}) must be satisfied
to solve the flatness problem. 
This constraint~(\ref{condi:flatness}) contains
the displacement of the center of the nucleated bubble from the center
of the symmetry of the bulk-brane system, $|\Delta|$,
whose dependence on the model parameters is not obvious.  
We shall construct a model in which the expectation value of $|\Delta|$ is 
small
and give a rough estimate for the expectation value of
$|\Delta|$ to obtain
the condition solely written in terms of the parameters of the model.

{}For this purpose, we examine
the process of the false vacuum decay with nucleation of a true vacuum
bubble on a space with boundary,
which is described by an instanton,
i.e., an Euclidean classical solution which
contains a time slice corresponding to the field configuration
at the instance of the nucleation~\cite{CD}.

As proven in Ref.~\cite{CGM}, the bubble nucleation
with the highest possible symmetry is the most favored tunneling
process in the case of the false vacuum decay
in Minkowski background {\em without boundaries}.
If the quantum tunneling were in general dominated by the process with 
the highest symmetry, the bubble nucleation with $|\Delta|=0$ 
would be the dominant process. 
Although this expectation is in favor of our attempt to construct
the model, we have to
be careful on the fact that in the system without boundaries, the
nucleated bubble possesses translational invariance. As a result,
the shifted bubble is as likely to nucleate
as the unshifted, contrary to the demand of our scenario that the
unshifted is to be the most likely.
Even for the system with a boundary,
we would expect the same trouble if there were
no interaction between the tunneling field and the boundary brane.
We therefore consider the otherwise.
Our model is
composed of a bulk scalar field $\phi$ minimally coupled to gravity
with its bulk potential $V(\phi)$ having minima at $\phi=\phi_T$ and
$\phi=\phi_F$, which correspond to a true vacuum and a false vacuum,
respectively, and the potential $U(\phi)$
localized on the 3-brane. This potential $U(\phi)$ introduces the degree
of freedom that controls the strength of the interaction
between the tunneling  field and the boundary brane.

Since our interest is on the bubble with its center in the vicinity of
the center of the bulk-brane system,
we consider an instanton with a bubble whose center
coincides with that of the bulk-brane system,
and treat the displacement of the bubble center
as quantum fluctuations around this instanton.
When the center of the nucleating bubble coincides with the center
of the symmetry of the bulk-brane system, the instanton has the same
$O(5)$-symmetry as 
that of the Euclideanized bulk-brane system.
Hence, the Euclideanized
geometry of the instanton may be described as
\begin{eqnarray}
   ds^2=dy^2 + \alpha^2(y) d\sigma^2_{S_4}\,, 
\label{metric}
\end{eqnarray}
where $d\sigma^2_{S_4}$ is the metric of unit four-sphere, and
the scalar field $\phi$ is a function of $y$.
The 3-brane is located at $y=y_{B}$, and thus $\alpha(y_B)=\alpha_B$ 
stands. 
To simplify our discussion, we neglect
the back reaction to the geometry due to the non-trivial scalar field
configuration. To be more precise, we discuss the nucleation of the
scalar field bubble on the fixed background with a constant curvature, 
whose curvature length $\ell$ is determined by the 
energy density in the true vacuum.
Since we are interested in the nucleation of an AdS-bubble, 
which yields a braneworld of the Randall-Sundrum type,
we consider the warp factor given by $\alpha(y)=\ell \sinh (y/\ell)$.
When we discuss the cases in which the state before transition
is de Sitter or Minkowski spacetime, the approximate treatment
fixing the background geometry may not be valid.
In such cases, we need to develop the formulation
taking into account not only the back reaction to the geometry
but also the effect of gravitational fluctuations, 
both of which we do not  discuss in the present paper.

The equation that the instanton $\bar\phi(y)$ obeys is
\begin{eqnarray}
  \bar\phi'' + 4\frac{\alpha'}{\alpha} \bar\phi'
     = \partial_{\bar\phi} V(\bar\phi)
       +\delta(y-y_B)\partial_{\bar\phi} U(\bar\phi)\,,
\label{fieldEq}
\end{eqnarray}
where {\em prime} denotes a differentiation with respect to $y$.
The condition that the instanton is regular at the origin
requires the boundary condition  $\bar{\phi}'(0)=0$. 
Integrating Eq.~(\ref{fieldEq}) in the vicinity of the brane and
recalling the ${\mathbb Z}_2$ symmetry across the brane, we see that
the boundary condition on $\bar\phi(y)$ at the brane is given by
\begin{eqnarray}
    2\bar{\phi}'(y_B)=-\ppp_{\phi}U(\bar\phi(y_B)) \,. 
\label{boundaryPhi}
\end{eqnarray}

Next, we derive the probability distribution
of the bubble nucleation as a function
of $\Delta$, by considering quantum
fluctuations around the instanton.
The Euclidean action is expanded to the second order with
respect to the fluctuation field $\delta\phi$ as
\begin{eqnarray}
   && S_E
       =S_E[\bar{\phi}]
         +{1\over 2}\int dy\, d\Omega\, \alpha^4\,  
           \delta\phi\, \hat {\cal O} \delta\phi +\cdots\, ,
\label{SeExpanded}
\end{eqnarray}
with
\begin{eqnarray}
    \hat{\cal O}:=-\frac{\partial^2}{\partial y^2} 
            - 4\frac{\alpha'}{\alpha}\frac{\partial}{\partial y} 
            - {{1} \over \alpha^{2}}\triangle_{S_{4}} 
            +\partial_\phi^2 V(\bar{\phi})
            +\delta(y-y_B)\partial_\phi^2 U(\bar{\phi}),
\end{eqnarray}
where $\triangle_{S_4}$ is the Laplacian operator on a unit four-sphere.
The exponentiated action $e^{-S_E}$ gives the probability of
the bubble nucleation process that the analytic continuation of 
$\bar{\phi}+ \delta\phi$ describes. Hence, if we choose $\delta\phi$
corresponding to the shift of the nucleated bubble,
we obtain the distribution function of $\Delta$.
The change in the field configuration after
an infinitesimal global translation by an amount of $\Delta=|\delta x^\mu|$ 
in proper distance is given by 
\begin{equation}
   \delta\phi=\bar\phi(y- y_{,\mu}\delta x^\mu)-\bar\phi(y)
   =\chi(y) Y_{1}(\Omega^\mu) \Delta, 
\end{equation}
where $\chi(y):= - \bar{\phi}'(y)$ and $Y_{{\bf k}}(\Omega^{\mu})$ is 
the normalized eigen functions of the Laplace operator 
$\triangle_{S_4}$ which satisfies
$[\Delta_{S_4}+k(k+3)] Y_{{\bf k}}=0$ and
$\int d\Omega\, Y_{\bf k}^2 =1$.
The substitution of
$\delta\phi=\chi(y)Y_{1}(\Omega^\mu)\Delta$ into the action does not
suffice for our goal. 
This naive substitution can underestimate
the amplitude of fluctuations of $\Delta$, since
$\chi(y)Y_{1}(\Omega^\mu)$ is not in general an eigen function of the
operator ${\cal O}$ satisfying the boundary condition
at the brane, as we shall see below.

To obtain the eigen function, we decompose the fluctuation field
for $k=1$ as 
$\delta\phi=-\varphi_1(y) Y_{1}(\Omega^\mu) \Delta$. 
Then the eigen value equation
$ \alpha^2 \hat{\cal O}\delta\phi=\lambda_1 \delta\phi$ with the lowest eigen
value $\lambda_1$ becomes
\begin{eqnarray}
       \left[-\frac{d^2}{dy^2}-4\frac{\alpha'}{\alpha}\frac{d}{dy}
            + { {4} \over \alpha^{2}}
            +\partial_\phi^2 V(\bar{\phi})
      \right]\varphi_1 (y)= \lambda_1  \alpha^{-2}\varphi_1 (y)  \,.
\label{eigenEq}
\end{eqnarray}
The boundary condition at the brane is given by
\begin{eqnarray}
    2\partial_y \delta\phi= - \delta\phi
       \ppp_{\phi}^2 U(\bar\phi(y_B))\,.
\label{varphi-bound}
\end{eqnarray}
Differentiating Eq.~(\ref{fieldEq}) with respect to $y$, one can verify
\begin{eqnarray}
        \left[-\frac{d^2}{dy^2}-4\frac{\alpha'}{\alpha}\frac{d}{dy} 
            +{4 \over \alpha^{2}}
            +\partial_\phi^2 V(\bar{\phi})
      \right]\chi(y)= 0 \,,
\label{chiEq}
\end{eqnarray}
which means $\chi(y)$ satisfies the eigen equation~(\ref{eigenEq}) for
zero eigen value in the bulk. 
However, $\chi(y)$ does not satisfy
the boundary condition (\ref{varphi-bound}) in general.
When we consider the thin-wall limit, for which $\chi(y)=0$ near 
$y=y_B$, 
with vanishing boundary potential, $U(\phi)=0$,
$\chi(y)$ does satisfy the boundary condition.
In this case, substituting
$\delta\phi=-\chi(y) Y_{1}(\Omega^\mu) \Delta$ into
Eq.~(\ref{SeExpanded}), we find that $S_E$ does not depend on $\Delta$.
This means that the bubble
nucleates with the same probability
independent of the location of the nucleation center
as in the case without the boundary brane.
The result is as expected since there is no interaction
between the tunneling field and the boundary brane in this case.

In general, since $\chi(y)$ does not satisfy
the boundary condition (\ref{varphi-bound}),
$\chi(y)Y_{1}(\Omega^\mu)$ is not the lowest $k=1$ eigen mode.
Nevertheless, unless extreme cases are concerned,
we can expect that the true lowest eigen value is still close to $0$.
If it is the case, the corresponding mode $\varphi_1(y)$ will also
be close to $\chi(y)$, at least for $y\lesssim y_W$.
Then,
by choosing the normalization of $\varphi_1$ such that they satisfy
$\lim_{y\rightarrow0}\varphi_1/\chi=1$,
we can still interpret $\varphi_1 Y_{1}(\Omega^\mu)$
as the mode that represents the shift of the nucleation center.

The fact that $\chi(y)$ does not give an eigen function
means that the most probable field configuration
for an off-centered transition is not a simple translation
of the symmetric instanton.
The field configuration must be {\em deformed}
due to the effect of interaction between the tunneling field
and the boundary brane.
This deformation results in non-vanishing $\lambda_1$.
If $\lambda_1$ is positive, the shift of the
center of the nucleating bubble is associated with the
deformation that increases $S_E$.
As a result, the probability of the nucleation of
the off-centered bubble is suppressed.

Substituting $\delta\phi=\varphi_1(y) Y_{1}(\Omega^\mu) \Delta$ into
Eq.~(\ref{SeExpanded}), we have
\begin{eqnarray}
   P(\Delta)\propto e^{-S_E}=e^{-S_E[\bar{\phi}]}
          \exp\left[
         -\frac12\left(\lambda_1\int_0^{y_B} dy\, \alpha^2 \varphi_1^2\right)
            \Delta^2\right]\,
\end{eqnarray}
from which we see that  the most of the bubbles nucleate with its center
displaced from
the center of the bulk-brane system less than the amount
\begin{equation}
 \sqrt{\langle \Delta^2\rangle} \approx  \left[\lambda_1
      \int_0^{y_B} dy\, \alpha^2 \varphi_1^2\right]^{-1/2}   
\label{amp} 
\end{equation}
for a positive value of $\lambda_1$, while a negative value of
$\lambda_1$ means that the off-centered bubble is more likely to
nucleate.

Now, what we need to evaluate is the quantity on the right hand side of
this equation. For this purpose, first we note that $\varphi_1$ is the
eigen function of Eq.~(\ref{eigenEq}) with the lowest eigen value and
the instanton solution $\bar\phi(y)$ is monotonic.  Thus, we can choose
both $\varphi_1$ and $\chi$ be non-negative without loss of generality.
Then, we consider the Wronskian relation
between $\varphi_1$ and $\chi$,
\begin{equation}
 \alpha^4\left[ \varphi_1(y) \chi'(y)- \varphi'_1(y)\chi(y)\right]
    =\lambda_1 \int_0^y dy\, \alpha^2 \chi \varphi_1\,,
\label{Wron}
\end{equation}
which follows from Eqs.~(\ref{eigenEq}) and (\ref{chiEq}).
When the potential of $\phi$ localized on the
brane $U(\phi)$ is absent,
we find $\chi(y_B)=0$ from Eq.~(\ref{boundaryPhi}).  
This implies $\chi'(y_B)<0$ because $\chi$ is non-negative.
In this case, the Wronskian relation reduces to
$\alpha^4(y_B) \varphi_1(y_B)\chi'(y_B)
 =\lambda_1\int_0^{y_B} dy \,\alpha^2\chi\varphi_1$.
The integral on the right hand side is always positive
while the product on the left hand side is non-positive. 
Thus, in this case $\lambda_1$ is non-positive.
This means that the nucleation of the off-centered bubble is not
suppressed but rather enhanced in the system with no localized potential
$U(\phi)$.

When $U(\phi)$ is present, from Eq.~(\ref{Wron}) we obtain 
\begin{equation}
  \lambda_1 =\left(\int_0^{y_B} dy\, \alpha^2 \chi \varphi_1\right)^{-1}
      \left[\alpha^4\chi \varphi_1
      \left({\chi'\over\chi}-{\varphi_{1}'\over\varphi_{1}}\right)
            \right]_{y=y_B}.
\label{Wron2}
\end{equation}
Since $\chi\varphi_1\ge0$,
the signature of the eigen value ${\lambda_1}$ is determined by
the factor
\begin{eqnarray}
 \nu:=
   \left[{\chi'\over\chi}-{\varphi_{1}'\over\varphi_{1}} \right]_{y=y_B}
   =-4 \frac{\alpha'(y_B)}{\alpha(y_B)}
            -2 \frac{\partial_{\phi} V(\bar\phi(y_B))}
                        {\partial_{\phi}U(\bar\phi(y_B))}
     + \frac{1}{2} \partial_{\phi}^2 U(\bar\phi(y_B))
        \,,  
\label{+veCond}
\end{eqnarray}
where we have used the boundary conditions (\ref{boundaryPhi})
and (\ref{varphi-bound}).

Now we present a rough estimate of the right hand side of
Eq.~(\ref{amp}) under the following assumptions: 
\begin{eqnarray}
  \begin{array}{cl}
  \mbox{(i)} &
  \mbox{The order of magnitude of $\varphi_1$ is not significantly different
        from that of $\chi$.}\\
  \mbox{(ii)} &
  \mbox{$\chi$ does not rapidly go to zero near the boundary $y=y_B$. 
        }
  \end{array}
\label{est-cond}
\end{eqnarray}
In the present very rough estimate, we do not care about a factor
of order 10 or $10^2$. 
\footnote{
People who are familiar with the ordinary false vacuum decay
without boundary may think that the second assumption is not appropriate. 
It is important to note here that we have the boundary potential
$U(\phi)$ in our present model, which does not exist in the ordinary
false vacuum decay.    
The minimum of this boundary potential
in general does not coincide with that of the bulk potential $V(\phi)$.
As a result, the tunneling field is still on the slope of the
bulk potential near the boundary.
}.

As a typical scale for $\chi^2$, we introduce an energy scale $M$, i.e,
$\chi^2=O(M^5)$.
This energy scale is also related to the difference of the vacuum energy
between true and false vacua. Integrating Eq.~(\ref{fieldEq}),
we obtain
\begin{equation}
 \delta V\approx  {1\over 2}\chi^2(y_B)
           +4\int_0^{y_B}\!\!dy {\alpha'\over \alpha} \chi^2
    \approx M^5 \,. 
\label{deltaV}
\end{equation}
Here we have assumed that $y_B$ is not extremely large compared to $\ell$. 

Under these assumptions, Eqs.~(\ref{amp}) and (\ref{Wron2})
give
\begin{eqnarray}
   \sqrt{\langle \Delta^2\rangle}
                       \approx \frac{1}{\alpha_B^{2}\sqrt{\nu M^5}} \,.
\label{est-xi2}
\end{eqnarray}

\section{Constraint on the model parameters}
\label{Sect:Constraint}

In this section we discuss the constraints 
on the model parameters. 
Here, we discuss the two different regimes respectively: 
case (a), $\ell_T^{-2}\gg\alpha_B^{-2}$ and
case (b), $\ell_T^{-2}\alt \alpha_B^{-2}$. 
With the condition for the cosmological constant $\Lambda_4$ 
to vanish (See Eq.~(\ref{ellT})), 
Eq.~(\ref{alphaB}) leads to 
\begin{eqnarray}
\alpha_B^{-2}=\ell_T^{-2}-\ell_F^{-2}. 
\label{alphaB2}
\end{eqnarray}
Using this relation, we will find that 
case (a) and case (b) defined here are equivalent to 
those defined in Sec.~\ref{SubSect:evolution}.

We discuss the constraints 
on the model parameters on the $m_5$-$M$ plane, 
where $m_5$ is the 5-dimension Planck mass 
defined by $\kappa_5=6/m_5^3$. 
First we identify the boundary of the two regimes 
defined above on this $m_5$-$M$ plane. 
On the one hand, the 4-dimensional Planck mass after the big-bang 
is related to $m_5$ and $\ell_T$ by 
\begin{equation}
 m_{pl}^2\approx m_5^3 \ell_T \,.
\label{ellT2}
\end{equation}
On the other hand, 
recalling that 
$\ell_T^{-2}-\ell_F^{-2}=\kappa_5 \delta V/6$, 
from Eqs.~(\ref{deltaV}) and (\ref{alphaB2}) we obtain 
\begin{equation}
\alpha_B^2\approx {m_5^3\over M^5} \,, 
\label{alpha-m5-M}
\end{equation}
Hence, we have 
\begin{eqnarray*}
M/m_{pl}\ll (m_5/m_{pl})^{1.8},\quad \mbox{for case (a)}; \quad
M/m_{pl}\agt (m_5/m_{pl})^{1.8},\quad\mbox{for case (b)}. 
\end{eqnarray*}

\svs 
\noindent\underline
{\large \it Flatness} : 
The resulting universe must be sufficiently 
flat. This condition gives the constraint~(\ref{condi:flatness}). 
For case (a) we have $\ell_T\approx \ell_F$ from Eq.~(\ref{alphaB2}). 
Then, using Eqs.~(\ref{ellT2}), (\ref{est-xi2}) and (\ref{alpha-m5-M}), 
we obtain 
\begin{eqnarray}
 && \frac{\nu}{M} 
    \agt 10^{64}\widetilde{C}_W
      \left(\frac{M}{m_{pl}}\right)^{13/2}
        \left(\frac{m_5}{m_{pl}}\right)^{-3/2}, 
 \quad \mbox{for case (a)}, \cr
  && \frac{\nu}{M} 
    \agt 10^{64}\widetilde{C}_W
      \left(\frac{M}{m_{pl}}\right)^{51/4}
        \left(\frac{m_5}{m_{pl}}\right)^{-51/4}, 
 \quad \mbox{for case (b)}, 
\label{off-center}
\end{eqnarray}
where $\widetilde{C}_W:=\sqrt{\alpha_W/\delta_W}$ and 
the parameter $\delta_W$ is defined by $\sigma_W=M^5\delta_W$.

\svs 
\noindent\underline
{\large \it Growing mode}\footnote{This constraint 
was absent in the earlier version of this paper. 
The importance of this constraint was pointed out 
by J. Garriga\cite{private}.} : 
For the bubble to collide with the brane, 
deviation of the nucleated bubble from the most symmetric instanton 
is necessary, 
and this deviation $\delta\phi$ needs to increase as 
the bubble expands. That is,
the existence of a growing mode of $\delta \phi$ is necessary.  
Otherwise, $\delta \phi$ oscillates 
with decreasing amplitude around the unperturbed trajectory 
of the most symmetric configuration, 
and thus the bubble never hits the brane.

Assuming that the $y$-dependence of the fluctuation 
$\delta \phi$ is identical to $\varphi_1(y)$ given in the preceding 
section, the evolution equation of $\delta \phi$, 
$[
         \Box_{(5)} - \partial^2_\phi V(\phi)    
   ]  \delta \phi = 0,  
$
reduces to
\begin{eqnarray}
   \left(\Box_{(4)} - \lambda_1 +4 \right) \delta \phi = 0, 
\label{eq:evolve-fluctuation}
\end{eqnarray}
where we have used  Eq.~(\ref{eigenEq}), and 
$\Box_{(5)}$ and $\Box_{(4)}$ are, respectively, 
the d'Alembertian on the $5$-dimensional background 
bulk with the metric (\ref{metric}) and that on the worldsheet 
of the unperturbed bubble wall, i.e., a $4$-dimensional de Sitter
space. 
Thus, the condition for $\delta\phi$ to have the growing mode is
\ben
    \lambda_1 \leq 4 \,. 
\label{condi:upperbound}
\een 

Using the estimations in Eqs.~(\ref{Wron2}) and (\ref{+veCond}), we find
$\lambda_1 \approx \alpha_B^2 \ell^{-1}_F \nu$ for case (a), 
and $\approx \alpha_B \nu$ for case (b).  
Thus, the condition (\ref{condi:upperbound}) becomes  	
\begin{eqnarray}
 && \frac{\nu}{M} 
     \lesssim \bigg(\frac{M}{m_{pl}} \bigg)^4
                    \bigg(\frac{m_5}{m_{pl}}\bigg)^{-6}\,, 
 \quad\quad \mbox{for case (a)}, \cr
 && \frac{\nu}{M} 
     \lesssim \bigg(\frac{M}{m_{pl}} \bigg)^{3/2}
                    \bigg(\frac{m_5}{m_{pl}}\bigg)^{-3/2}\,, 
 \quad \mbox{for case (b)}.  
\label{grow}
\end{eqnarray}

Parameter $\nu$ represents the strength of 
interaction between the brane and the bubble. 
The conditions (\ref{off-center}) and (\ref{grow}) 
indicate that the interaction should be strong enough
to force the bubble to nucleate near the center of the 
brane-bulk system but weak enough to allow the bubble fluctuation to grow. 
Both conditions can be satisfied simultaneously for  
\begin{eqnarray}
 && \frac{M}{m_{pl}}
    \lesssim 10^{-25.6}\widetilde{C}_W^{-2/5}
      \bigg(\frac{m_5}{m_{pl}}\bigg)^{-1.8}, 
 \quad \mbox{for case (a)}, \cr 
 && \frac{M}{m_{pl}}
    \lesssim 10^{-5.7}\widetilde{C}_W^{-4/45}
       \frac{m_5}{m_{pl}}\quad, 
 \quad\quad \mbox{for case (b)}, 
\label{collide}
\end{eqnarray}
with an appropriate choice of the parameter $\nu$. 
We discuss the issue of tuning $\nu$ later. 

\svs 
\noindent\underline
{\large \it Reheating temperature} :
In order that the standard nucleosynthesis scenario works, 
the reheating temperature must be higher than, say, about 10 MeV. 
Using the estimate (\ref{rhoBB}), this condition gives a constraint 
\begin{eqnarray}
   {M\over m_{pl}}\gtrsim 10^{-17}\widetilde{C}_W^{2/5} 
 \left({m_5\over m_{pl}}\right)^{0.6}\,, 
  \quad \mbox{for case (a)}, \cr 
   {M\over m_{pl}}\gtrsim 10^{-34}\widetilde{C}_W^{4/5} 
 \left({m_5\over m_{pl}}\right)^{-0.6}\,, 
  \quad \mbox{for case (b)}.  
 \label{reheating}
\end{eqnarray}

\svs 
\noindent\underline
{\large \it Newton's law} :
It is also important to quote the well-known constraint on $m_5$ 
which is common to all the braneworld models of the Randall-Sundrum type. 
It is usually stated as a constraint on $\ell_T$ 
as $\ell_T \lesssim 1$mm, which is necessarily satisfied for 
the gravity induced on the brane to reproduce the 
Newton's law at the scale longer than about 1mm.  
This can be recast into a constraint on $m_5$ as 
\begin{equation}
  {m_5\over m_{pl}}\gtrsim 10^{-11} \,. 
\label{Newton'sLaw}
\end{equation}

\svs 
\noindent\underline
{\large \it 5D quantum effect}:
If $\ell_T$ becomes smaller than the 5-dimensional Planck length 
$m_5^{-1}$, the classical treatment of the bulk will not be 
justified. Hence, from $\ell_T \approx m_{pl}^2/m_5^3$, 
the parameter region which we can discuss is restricted to 
\begin{equation}
 {m_5\over m_{pl}}<1. 
\end{equation}

\begin{figure}  
  \begin{center}
    \begin{minipage}{7cm}
  \centerline{\epsfxsize = 7cm \epsfbox{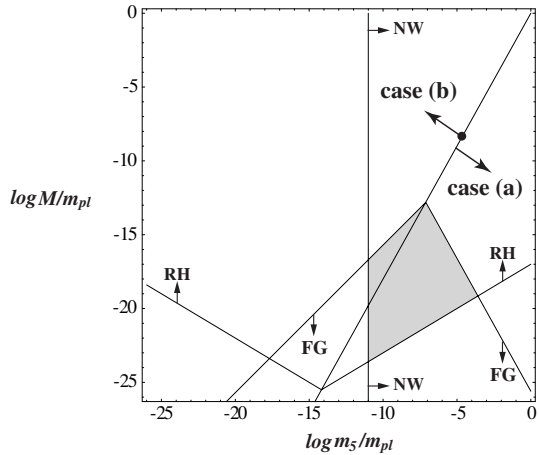}} 
    \end{minipage}
  \end{center}
\begin{center}
\begin{minipage}{14cm}
        \caption{\small   
                  The constraints on the parameters $(m_5,M)$ are
                  shown in the units of $m_{pl}$. The parameters in 
                  the shaded region are allowed for 
 $O(\nu/\ell_F^{-1})\approx 1$.
                  {\bf FG} denotes the constraint that the bubble
 collide with the brane and result in the sufficiently flat universe,
 as in Eq.~(\ref{collide}); and {\bf NW} denotes the
 constraint that the Newton's' law be valid on scale larger than 1mm, as
 in Eq.~(\ref{Newton'sLaw}). 
 {\bf RH} is the constraint on the reheating temperature 
 in Eq.~(\ref{reheating}). 
                 }  
 \protect  \label{fig:Mconstraint} 
\end{minipage}
\end{center}  
\end{figure}

All the constraints mentioned above are summarized in 
Fig.~\ref{fig:Mconstraint}. 
The allowed region of parameters looks wide. 
However, one must be careful about the interpretation 
of this figure because we have assumed tuning of the 
parameter $\nu$. We can estimate the natural order 
of magnitude of $\nu$ from the expression (\ref{+veCond}). 
The negative first term on the right hand side is 
estimated as follows. Suppose that the false vacuum 
is AdS. Then $\alpha(y)=\ell_F\sinh(y/\ell_F)$, and 
hence the first term on the right hand side of Eq.~(\ref{+veCond}) 
is estimated as 
$-4\ell_F^{-1}\coth (y_B/\ell_F) = 
-4\ell_F^{-1} (1+\ell_F^2\alpha_B^{-2} )^{-1/2} = -4\ell_T^{-1}$. 
As a naive expectation, the second and the third terms 
will be $O(M)$.
If we adopt this estimate, 
the requirement for the positivity of $\nu$ becomes 
\begin{eqnarray}
 {M\over m_{pl}}\agt \left({m_5\over m_{pl}}\right)^3.   
\end{eqnarray}

If we naively assume that $\nu$ is $O(M)$, the condition 
for the existence of a growing mode becomes 
\begin{eqnarray}
  &&   {M\over m_{pl}}\agt \left({m_5\over m_{pl}}\right)^{1.5}, 
      \quad \mbox{for case (a)}, \cr    
  &&   {M\over m_{pl}}\agt {m_5\over m_{pl}}\quad,\quad 
      \quad \mbox{for case (b)}.    
\end{eqnarray}
Hence, we find that there is no solution. 
However, it is possible to tune the parameter $\nu$ to be 
sufficiently small when $\ell_T^{-1}\approx M$, i.e, when
${M/m_{pl}}\approx (m_5/m_{pl})^{3}$. 
In this case, even if the second and third terms 
on the right hand side of Eq.~(\ref{+veCond})
are $O(M)$, they can cancel with the first term. 
Another possibility is to assume a cancellation 
between the second and the third terms. 
If there is a mechanism to tune $\nu$ as small 
as the amplitude of the first term, 
the condition for the existence of a growing mode 
becomes 
\begin{eqnarray}
  &&   {M\over m_{pl}}\agt \left({m_5\over m_{pl}}\right)^{1.5}, 
      \quad \mbox{for case (a)}, \cr    
  &&   {M\over m_{pl}}\agt \left({m_5\over m_{pl}}\right)^{1.8}, 
      \quad \mbox{for case (b)}.
\end{eqnarray}
Hence, there is no parameter region which works for this choice of 
$\nu$ for case (a), while the allowed region shown in 
Fig.~\ref{fig:Mconstraint} is not reduced by imposing the condition 
$\nu\agt \ell_T^{-1}$ for case (b).

One may worry about the limitation of the validity of 
our current treatment of neglecting the gravitational back reaction 
in deriving the estimate (\ref{amp}). 
The condition for weak back reaction will be given by 
$1\gg (K^+-K^-)/K\approx
\kappa \sigma_W (\alpha_W^{-2}+\ell_F^{-2})^{-1/2}$.  
Using Eqs.~(\ref{alpha}) and (\ref{alphaB2}), we have 
$ 
 \kappa \sigma_W (\alpha_W^{-2}+\ell_F^{-2})^{-1/2}  
 \approx 6 \kappa^2 \sigma_W^2 \alpha_B^2  
         (9 - \kappa^2 \sigma_W^2 \alpha_B^2)^{-1}  
$. 
Hence the weak backreaction condition 
is reduced to $\kappa_5\sigma_W\alpha_B\ll 1$. 
Since the width of wall $\delta_W$ 
must be at most of $O(\alpha_W)$, we have 
$\sigma_W=\delta_W M^5\ll \alpha_B M^5$. 
As long as we use this estimate, the condition for 
the weak back reaction 
$\kappa_5\sigma_W\alpha_B\ll 1$ is always satisfied.

\section{Summary and Discussion} 
\label{Sect:Conclusion} 

We proposed a new scenario of inflation based on the 
Randall-Sundrum braneworld model~\cite{RS}. 
The model consists of 5-dimensional bulk  
with ${\mathbb Z}_2$-symmetry and a brane 
which is located at the fixed point.  
In our model, when the universe first nucleates, 
the brane is in de Sitter expansion phase as in the scenario of 
the braneworld creation discussed in Ref.~\cite{Creation}. 
In addition to it, the 5-dimensional bulk is assumed to be in a false 
vacuum initially.  
Then, eventually false vacuum decay will occur through nucleation of a 
true vacuum bubble. This nucleated bubble expands and hits the brane. 
Then the energy of the bubble wall is converted into the 
thermal energy of the fields living on the brane. 
Then the brane-bubble collision results in the creation of 
a hot big bang universe. 
After the transition, the energy density in the bulk decreases. 
As a result the effective cosmological constant induced on the brane 
also decreases. Thus the inflationary phase terminates. 
Although the usual fine-tuning problem still remains, it is possible 
to choose the model parameters so that the effective 
cosmological constant vanishes.

In this scenario the spatial curvature radius 
at the moment of the big bang is related to 
the location of the bubble nucleation.  
We found that, to solve the flatness problem, the bubble must nucleate 
with its center being very close to the center of the symmetry 
of the bulk-brane system. 
Under the assumption that the effect of the 
gravitational back reaction is small, we derived an estimate 
for this probability distribution.  
The condition for the weak back reaction was shown to hold 
when the curvature radius of the bubble wall is 
much smaller than that of the brane. 
Then, we found that 
this required concentration of the nucleation point can  
be realized if the tunneling field has an appropriate potential 
localized on the brane.  
Another important constraint comes from the condition that 
fluctuations of the nucleated bubble wall continue to 
grow. As mentioned above, due to the requirement of solving 
the flatness problem it was necessary to introduce an interaction 
between the brane and the bulk field.  
This interaction inevitably increases the effective 
mass squared of the perturbation modes corresponding to wall fluctuations.
When there is no interaction, this effective mass squared is 
negative, which means that the wall fluctuation grows until 
it hits the brane. If this mass squared becomes positive, 
the fluctuation modes are stabilized, and the bubble wall 
never hit the brane. Hence, 
the interaction should be strong enough
to force the bubble nucleate near the center of the 
brane-bulk system but weak enough to let the bubble fluctuation grow.

We also discussed other two constraints on the model parameters.  
One comes from the condition that the reheating temperature 
is sufficiently high. 
The other is the condition that the deviation from the Newton's law 
is not allowed at the scale longer than 1mm. 
These constraints are summarized in Fig.~\ref{fig:Mconstraint}. 
Still a wide region in the parameter space is not excluded. 
However, the result must interpreted carefully. 
The interaction strength mentioned above needs 
to be tuned additionary. Unfortunately we could not find a 
natural explanation for this parameter tuning within our simple model.  
We expect future new invention on this point. 

Here we note that the geometry of our model has a lot of variety. 
The allowed region is divided into three portions as shown in 
Fig.~\ref{fig:Mconstraint}. The initial state of the 
bulk is anti-de Sitter spacetime for case (a), 
nearly flat spacetime for case (b$_1$), and 
de Sitter spacetime for case (b$_2$).

Here we wish to stress one distinctive feature of 
our model. 
We know that it is usually convenient to introduce a corresponding 
effective 4-dimensional picture even when we are interested in 
a 5-dimensional model.   
However, if we interpret the scenario proposed in this paper 
on the viewpoint of the 4-dimensional effective theory, 
it does not look quite natural. The phase transition 
occurs beyond the horizon scale in a completely synchronized manner. 
It will be necessary to consider a slightly complicated situation 
in order to explain such a process without assuming the 
existence of extra-dimension(s). 
This means that our scenario gives a new paradigm
opened for the first time in the context of the braneworld. 

Our scenario has the common idea with the interesting model 
proposed by Bucher\cite{Bucher} in the sense that the bubble 
collision in higher dimensional spacetime brings a big-bang 
to our 4-dimensional world realized on the brane. 
Although the basic idea is quite similar, these two models have 
many different aspects. 
In the model proposed by Bucher, a sufficiently 
big bulk is supposed to exist as an initial condition. 
The big-bang occurs through the collision 
of two nucleated bubbles. The flatness problem is solved by the 
large separation between the nucleation centers of the two bubbles. 
The place of collision is not special at all before the 
collision occurs. Hence, the tension of the brane formed 
after the bubble collision is brought by the colliding bubbles.  
On the other hand, in our present model the universe starts with 
a small bulk initially. 
As a consequence of this, the localization of the nucleation center of the 
colliding bubble became necessary instead of the large separation of 
the bubbles. In the present model, the brane with positive tension exists 
as a target for the bubble wall to collide. Hence, we could assume
that the tension of the brane is unchanged between before and after 
the bubble collision as one of the simplest possibilities. 

In the sense that our universe is realized inside a single nucleated 
bubble, our new scenario has a common feature with 
the one-bubble open inflation~\cite{oboi}, too. 
However, in the one-bubble open inflation, 
a slow roll inflation inside the nucleated bubble was necessary. 
Since the bubble wall, namely, the boundary surface of the old 
inflationary phase, is timelike, the universe inside the nucleated bubble 
is curvature dominant from the beginning. 
Hence the flatness problem is not solved 
without the second inflationary epoch. 
On the other hand, in the present scenario, 
the bubble nucleation occurs not on the brane but inside the bulk. 
The boundary surface of the inflation on the brane is spacelike, 
hence the brane universe can be sufficiently flat 
without introducing a second inflation.

A few comments on the assumptions we used are in order. 
In the process of the brane-bubble collision, 
we simply assumed that 
the energy of the wall is completely converted into the energy of radiation 
on the brane. 
Although so far there have appeared several works on the collision 
of thin shells~\cite{collision}, 
the mechanism of energy transfer at the collision of branes 
has not been made clear yet. 
Moreover, the process of collision will be heavily model dependent.
Hence, it is easy to imagine other possibilities. 
For example, 
the collision might be elastic, and then the bubble bounces 
into the bulk.  
The energy of the bubble wall may 
completely dissipate into the bulk as radiation.  
The energy dissipating into the bulk may produce the Weyl 
components of the bulk gravity. 
Then, they affect the evolution of the FLRW brane as dark radiation,  
whose energy density must be suppressed compared to that of 
the fields localized on the brane. 
Otherwise, the standard big-bang nucleosynthesis would not work.  
To avoid this problem, 
our scenario may require
a certain mechanism which realizes the efficient energy conversion 
from the bubble wall to the fields localized on the brane. 
This might be possible, 
for example, if a sufficiently large number of light fields 
which couple to the bulk scalar field $\phi$ reside on the brane. 
If the inverse of the bulk curvature radius $\ell_T^{-1}$ is larger 
than the reheating temperature, most of the KK modes of 
the bulk fields will not be excited by the brane-bubble collision.
Then, the number of 
relevant degrees of freedom localized on the brane is larger than 
that living in the bulk. In such a situation,  
once the equi-partition among these relevant degrees of freedom 
is established, 
the relative contribution from dark radiation is suppressed. 
Alternatively, we can construct a model in which 
such a relic dark radiation is diluted by 
a fairly short period inflation like thermal inflation~\cite{ThermalInflation} 
implemented by the potential $U(\phi)$ on the brane.

We also assumed that the brane tension $\sigma_B$ does not 
change before and after the collision.  
One may think it more natural that the brane tension $\sigma_B$ 
changes during the collision. 
If the change of $\sigma_B$ is 
hierarchically small compared to the value of $\sigma_B$, 
the present scenario works as in the same manner. 
To further investigate this issue, 
one needs to specify the details of the model,    
which will be supplied once we can embed this scenario in 
more fundamental theories such as string theory or M-theory.

We considered a single bubble nucleation in the bulk. 
There is, however, a possibility that many bubbles nucleate. 
Since we must take into account the interaction among the bubbles,  
the localization of the nucleation center of a single bubble
does not imply that of multiple bubbles. 
Suppose that this localization is achieved. 
Then, the situation will be similar to the single bubble case. 
Since vacuum bubbles expand exponentially,  
they immediately collide with each other after the nucleation 
and continue to expand as a single bubble. 
The bubble collision may produce inhomogeneities on the bubble wall. 
However the rapid expansion of the bubble wall will erase 
such inhomogeneities by the time of the brane big-bang. 
If the bulk field potential $V(\phi)$ has a number of different vacua and 
bubbles are nucleated in the different vacuum phases, the collision of 
the nucleated bubbles may produce topological defects of lower dimension, 
which could remain on the FLRW brane after the brane big-bang. 
However, provided again that the nucleation of bubbles is confined very near 
the center, one can expect that the abundance of such lower dimensional 
topological defects in a horizon scale of the FLRW brane is reduced 
sufficiently by the de Sitter-like expansion of the bubble wall, 
as the standard inflation solves the problem of unwanted relics. 

In the present paper, we did not estimate the expected amplitude 
of density fluctuations in this new model. 
The analysis of density perturbations for our present model 
will bring another meaningful constraint on the model parameters.  
In a slightly simpler setup corresponding to Bucher's model\cite{Bucher}, 
the density fluctuations due to the quantum fluctuations 
of the colliding bubble walls have been calculated in
Ref.\cite{JB,GT2} 
assuming that the self gravity of the bubble walls is weak. 
These analyses have shown that 
the scale-invariant spectrum of the density fluctuations does not 
easily arise in the Bucher's model by this mechanism.  
However, this fact does not exclude the possibility of this 
model because the amplitude due to the effect of the bubble 
wall fluctuations tends to be very tiny. 
To apply a similar analysis to our present model, further extension 
of the formalism will be necessary.

\section*{\bf Acknowledgments}

We would like to thank H. Kodama, K. Kohri and K. Nakao 
for comments and discussion. We would like to express special thanks to 
J. Garriga, who suggested several important improvements\cite{private}. 
A.I. and U.G. are supported by Japan Society for the Promotion of Science.  
The work by T.T. is supported in part by 
Monbukagakusho Grant-in-Aid No.~1270154 and Yamada foundation. 
The authors thank the Yukawa Institute for Theoretical Physics 
at Kyoto University. Discussions during the YITP workshop YITP-W-01-15 
on ``Braneworld - Dynamics of spacetime with boundary'' 
were also very useful to complete this work.

\appendix
\section*{Appendix: Derivation of Lorentz factor} 
\label{Lorentz-factor} 

In this appendix
we derive the Lorentz factor $\gamma$ of the bubble wall 
seen from an observer on the brane. 
The discussion in this appendix is based on the note by 
J. Garriga\cite{private}. 
  
We consider the case that 
the false vacuum bulk is anti-de Sitter space and 
the collision is of the type~(i), 
for which the intersection of the brane and the bubble is ${\Bbb H}^3$. 
When we consider the case of de Sitter false vacuum bulk, 
all we have to do is to replace 
\ben
  X^5 \to i X^5_{deS} \,, \quad 
  \ell_F \to i \ell_F^{deS} \,, \quad 
  \theta_1 \to i \theta_1^{deS} \,. 
\een 
Because of the symmetry, we would be able to 
concentrate on the case with  $X^2=X^3=X^4=0$ 
without loss of generality.
  
Since the brane trajectory $\Sigma_B$ is 
the intersection of the pseudo-sphere~(\ref{embed:deSitter}) and the plane 
$X^5 = \ell_F \beta_B$, any point $P = (X^0_P,X^1_P,0,0,0,X^5_P)$ 
on $\Sigma_B$ can be expressed in terms of a vector in ${\Bbb E}^{4,2}$ as 
\ben
   P = \alpha_B \sinh \phi_B {\partial \over \partial X^0} 
        + \alpha_B \cosh \phi_B {\partial \over \partial X^1} 
        + \ell_F \beta_B {\partial \over \partial X^5} \,. 
\een
This is simply obtained by Lorentz boost of a vector 
$ P_0 = \alpha_B ({\partial /\partial X^1}) 
         + \ell_F \beta_B ({\partial /\partial X^5}) 
$, 
with an angle $\phi_B$. 
Then a unit tangent vector $T_B$ to $\Sigma_B$ at $P$ 
can be obtained from its $\phi_B$-derivative as
\ben
    T_B = \cosh \phi_B {\partial \over \partial X^0} 
          + \sinh \phi_B {\partial \over \partial X^1} \,. 
\een 

On the other hand, the bubble wall trajectory $\Sigma_W$ is given 
as the intersection of the pseudo-sphere~(\ref{embed:deSitter}) and 
the plane defined by 
$
   \bar{X}^5 = \ell_F \beta_W \,, 
$ 
where the coordinates $\{ \bar{X}^M \}$ are related to the original 
coordinates $\{ \bar{X}^M \}$ by (\ref{ct1}).  

Then any point $Q = (X^0_Q,X^1_Q,0,0,0,X^5_Q)$ on $\Sigma_W$ can 
be expressed as a vector 
\ben
    Q = \alpha_W \sinh \phi_W {\partial \over \partial \bar{X}^0} 
        + \alpha_W \cosh \phi_W {\partial \over \partial \bar{X}^1} 
        + \ell_F \beta_W {\partial \over \partial \bar{X}^5} \,.  
\een 
In the original coordinates $Q$ is written as 
\bena
    Q &=& \alpha_W \sinh \phi_W {\partial \over \partial {X}^0} 
        + \left( 
                \alpha_W \cosh \phi_W \cosh \theta_1 
                 + \ell_F \beta_W \sinh \theta_1 
          \right) 
          {\partial \over \partial {X}^1} 
\nonumber \\
      &{}& 
         + \left( 
                \alpha_W \cosh \phi_W \sinh \theta_1 
                 + \ell_F \beta_W \cosh \theta_1 
          \right) 
          {\partial \over \partial {X}^5} \,. 
\eena 
Then, as in the case of $T_B$, a unit tangent vector $T_W$ to $\Sigma_W$ 
at $Q$ is obtained as 
\bena
   T_W =
             \cosh \phi_W {\partial \over \partial {X}^0} 
           + \sinh \phi_W \cosh \theta_1 {\partial \over \partial {X}^1} 
           + \sinh \phi_W \sinh \theta_1 {\partial \over \partial {X}^5} \,. 
\eena  

In order to obtain the Lorentz factor $\gamma$, we need to consider 
the two tangent vectors $T_B$ and $T_W$ at the same point on 
the intersection $\Sigma_B \cap \Sigma_W$. So, setting $P = Q$, we obtain
\bena  
  && \alpha_B \sinh \phi_B = \alpha_W \sinh \phi_W \,, 
\\
  && \cosh \phi_B 
 = \frac{\ell_F \beta_B \cosh \theta_1 
   - \ell_F \beta_W}{\alpha_B \sinh \theta_1}, 
\\
  && \cosh \phi_W  = 
\frac{\ell_F \beta_B - \ell_F \beta_W \cosh \theta_1}{\alpha_W \sinh \theta_1} 
\,. 
\eena 
Using these relations, the Lorentz factor is evaluated as 
\bena
 \gamma &:=& - \langle T_B, T_W\rangle 
   = \cosh \phi_B \cosh \phi_B 
                          - \sinh \phi_B \sinh \phi_W \cosh \theta_1 
\nonumber \\  
 &=& { \ell_F^2 (\beta_B \beta_W - \cosh \theta_1) \over \alpha_B \alpha_W} \,.
\eena

\end{document}